\documentclass[12pt]{article}
\usepackage{graphicx}
\usepackage{epstopdf}
\usepackage{amssymb,amsmath,epsfig}
\usepackage{fancyhdr}
\usepackage{titlesec}
\usepackage{lipsum} 

\begin{document}

\title {\bf Anisotropic Compact Star with Vaidya-Tikekar Potential in $f(Q)$ Gravity}

\author {M. Awais \thanks{mawaisarain181@gmail.com, muhammadawais.maths@gmail.com}
and M. Azam \thanks{azammath@gmail.com, azam.math@ue.edu.pk},
\\
Department of Mathematics,\\ Division of Science and Technology,\\ University of Education, Lahore, Pakistan.}

\date{}

\maketitle
\begin{abstract}
The goal of this work is to develop physical models for spherically
symmetric systems in the realm of $f(Q)$ gravity. The field equations
are set up for anisotropic fluid and formulate these equations using
physical ansatz of Vaidya-Tikekar solution. The intrinsic constants
of the model be ascertained by matching the interior solution to the
external Schwarzschild solution. The use of accessibility conditions
helps to understand the physical viability of the obtained model.
LMC X-4 pulsar is used as a toy model to test the viability of the model
and Mathematica software is used for deeper insights and
visualization of the physical parameters against fixed values of constants.
\end{abstract}

\section{Introduction}

Our knowledge of gravity and the nature of space-time was completely transformed
by Albert Einstein's $1915$ Theory of General Relativity (GR), which went on to become
a cornerstone of contemporary physics. With regard to Einstein's theory, mass and
energy warp space-time itself, offering a geometric view that differs from Newton's
classical theory which defined gravity as a force between masses. The gravitational
attraction that humans sense is created by this curvature. According to GR,
gravity is a direct result of the twisted space-time surrounding enormous
objects rather than a force that exists in space-time itself. The equivalence principle,
which holds that the effects of acceleration and gravity are equivalent locally, is
fundamental to this theory. Rather than being a conventional force like electromagnetism,
this principle emphasizes Einstein's deep realization that gravity is an intrinsic element
of the geometry of space-time itself. Although GR has been highly
successful, it encounters difficulties in situations involving extremely high energies and
very small distances, like those found near black hole (BH) singularities or in the early moments
of the universe. In these extreme conditions, combining GR with quantum mechanics has proven
challenging, leading to the investigation of alternative theories of gravity.

The recent developments in extending GR and alternate theories have seen the rise of various
theories like $f(R)$, $f(R, T)$, and others, which incorporate geometrical concepts like
torsion $T$ and curvature $R$. One such alternate is the Teleparallel equivalent of GR
(TEGR) \cite{1} - \cite{6}, where both curvature and non-metricity are absent. Another approach,
known as Symmetric Teleparallel (STEGR) equivalent of GR, is also equivalent to GR, but in
this case, both curvature and torsion vanish, with non-metricity $Q$ serving as an
alternative description of gravitation. This has led to the development of $f(Q)$ gravity
theory \cite{7} - \cite{21}. $f(Q)$ gravity is an alternative to GR that
replaces the traditional gravitational field equations with a function of $Q$. In this
framework, the non-metricity scalar $Q$, which is devoid of both curvature and torsion,
plays a central role in describing gravitational interactions \cite{22}. Here curvature
plays the function of non-metricity in $f(Q)$ gravity, as opposed to GR, where curvature
is a feature linked with a particular connection. Its significance is highlighted by the
identification of the Einstein pseudo-tensor as the non-metricity $f(Q)$, which in geometric
situations usually turns into a true tensor. As demonstrated by Jimenez et al. \cite{23},
this conceptual change has converted STG into coincident GR, or
$f(Q)$ gravity. $f(Q)$ gravity follows standard interpretations on a cosmological scale,
especially in the setting of the Friedmann-Lemaitre-Robertson-Walker (FLRW) universe. On
the other hand, cosmological perturbation theory reveals differences and several aspects of
cosmology have been explored (reference therein \cite{24,25}).

Junior et al. \cite{26} investigate new BH and black-bounce solutions
in $f(Q)$ gravity, with a focus on space-time regularity, geodesic extendability, and energy
requirements. They focus on how these energy conditions behave in black-bounce circumstances.
Parsaei et al. \cite{27} investigate exact asymptotically flat wormhole (WH) solutions in
STEGR. They investigate WH solutions using modified field equations and various equations of state,
comparing the shape function and energy conditions for different $f(Q)$ models. Banerjee et al. \cite{28}
investigate WH models in the $f(Q)$ gravity framework. Based on recent research \cite{29,30},
they analyze field equations for spherically symmetric topologies, energy conditions and identify
differences between them. Morover, the WH solutions of static, spherically
symmetric studied in \cite{31} and concluded that null energy requirements broken near the WH's throat.
Their findings show that WH solutions are not viable for the quadratic form of function $f(Q)$.
Hassan et al. \cite{32} investigate the impact of the Generalized Uncertainty Principle
(GUP) parameter on Casimir WH in $f(Q)$ gravity, specifically how $Q$ affects gravitational forces.
They investigate WH solutions under anisotropic fluid conditions, violation of null energy
conditions (NEC) at the neck, and state equation parameters
under GUP effects.

Research on compact stars and BHs with $f(Q)$ gravity has revealed different features,
such as energy requirements, geodesic paths, and stability \cite{33} - \cite{36}. Studies on compact stars
\cite{37} - \cite{39} concluded that how $f(Q)$ gravity influences their equilibrium, mass-radius
relations, and high-density matter behavior. Similarly, various studies on black holes \cite{40,41} has concentrated
on the impact of $f(Q)$ gravity on horizons and thermodynamics. These studies hope to improve our
understanding of high-density astrophysical objects and the physics of gravity. Recently, we have formulated
a viable solution in $f(Q)$ gravity with Tolman-IV ansatz and test its viability for pulsar LMC X-4
\cite{42}. Considerable work has been done on the formation of primordial BHs in the context
of $f(Q)$ gravity \cite{43,44}.

Using the $f(Q)$ framework, this research explores the characteristics of spherically
symmetric systems with anisotropic matter distribution. The paper is set up on the following
pattern. A short overview of basics of $f(Q)$ gravity and fundamental equations used in the manuscript
is provided in Section \textbf{2}. In section \textbf{3}, an ansatz, Vidya-Tikekar is
employed to the field equations and match the internal
solution with external vacuum solution by assuming zero pressure at the boundary
to get expressions for constant in term of physical parameters, i.e., mass and radius of a star.
A detailed physical study of the model is provided in Section \textbf{4},
where we compute and graph the star's various physical parameters to ensure its stability.
In Section \textbf{5}, TOV equation is utilized to ensure model's stability.
Summery is presented in section \textbf{6}.

\section{$f(Q)$ Gravity and Field Equations}

{\allowdisplaybreaks
In this section, a brief introduction of $f(Q)$ gravity is given. For this, consider
the action as\cite{45}
\begin{equation} \label{1}
 S = \int \sqrt{-g}\; d^{4} x ~[ \frac{1}{2} ~ \textit{f} (Q) + L_{m}],
\end{equation}
where $Q$ is the non-metricity scalar and $g$ is the determinant.
The Lagrangian density of matter is represented by $L_{m}$.
Furthermore, the non-metricity tensor is defined by
\begin{equation}\label{2}
 Q_{\beta\mu\tau} \equiv \bigtriangledown_{\beta} g_{\mu\tau}.
 \end{equation}
The standard form of the affine connection, which is composed of
three different parts given by
\begin{equation}\label{3}
 \Gamma_{\mu\tau}^{\beta} = \lbrace_{\mu\tau}^{\beta}\rbrace + K_{\mu\tau}^{\beta} + L_{\mu\tau}^{\beta},
 \end{equation}
the Levi-Civita connection, $\lbrace_{\mu\tau}^{\beta}\rbrace$, and have the form
\begin{equation}\label{4}
 \lbrace_{\mu\tau}^{\beta}\rbrace = \dfrac{1}{2} g^{\beta\chi}(\partial_{\mu} g_{\chi\tau} +
 \partial_{\tau} g_{\chi\mu} - \partial_{\chi} g_{\mu\tau}).
 \end{equation}
The term $ K_{\mu\tau}^{\beta} $ in the given expression indicates the contortion
\begin{equation}\label{5}
 K_{\mu\tau}^{\beta} = \dfrac{1}{2} T_ {\mu\tau}^{\beta} + T_ ({\mu \;^{\beta}\; \tau\;_{)}},
 \end{equation}
where the torsion tensor, $ T_ {\mu\tau}^{\beta} \equiv 2 \;
\Gamma_{\mu\tau}^{\beta} $,  is specified as the antisymmetric component of the affine
connection, whereas, the disformation tensor $ L_ {\mu\tau}^{\beta} $
has the form
\begin{equation}\label{6}
 L_ {\mu\tau}^{\beta} = \dfrac{1}{2} Q_ {\mu\tau}^{\beta} - Q_ ({ \mu \;^{\beta}\; \tau\;_{)}}.
 \end{equation}
The non-metricity conjugate is computed as
\begin{equation}\label{7}
 P_{\gamma\tau}^{\lambda} = - \dfrac{1}{2} L_ {\gamma\tau}^{\lambda} - \dfrac{1}{4}
 [(Q'\;^{\lambda} - Q^{\lambda}) g_{\gamma\tau} + \delta^{\lambda} (_{\gamma} Q _{\tau}) ],
 \end{equation}
with two different traces
$$
 Q_{\lambda} = Q_{\lambda}\;^{\gamma}\;_{\gamma}   \; \;  ; \; \; Q'_{\lambda} = Q^{\gamma}\;_{\lambda \; \gamma} $$
which can be rewritten as $$  Q_{\lambda} = g^{\gamma\tau}\; Q_{\lambda\gamma\tau} \; \;;\; \; Q'_{\lambda} = g^{\gamma\tau} \; Q_{\gamma\lambda\tau}.$$
The subsequent field equations can be obtained by varying the action given in Equation (\ref{1})
corresponding to the metric $ g_{\mu\tau}$
\begin{equation}\label{9}
- 8 \pi T_{\mu\tau} = \dfrac{2}{\sqrt{-g}} \bigtriangledown_{\beta} (\sqrt{-g} \;f_{Q}\; P^{\lambda}_{\mu\tau}) + \dfrac{1}{2}\; g_{\mu\tau}\; f + f_{Q}\;(P_{\mu\lambda \chi}\;Q_{\tau}\;^{\beta \chi} - 2 Q_{\beta \chi\mu}\;P^{\beta \chi}\;_{\tau}),
\end{equation}
where $ f_{Q} \equiv \partial_{Q} f(Q)$ and the standard representation of the energy-momentum tensor is
defined by
\begin{equation}\label{10}
T_{\mu\tau} \equiv - \dfrac{2}{\sqrt{-g}} \dfrac{\delta\;(\sqrt{-g} \; L_{m})}{\delta g^{\mu\tau}}.
\end{equation}
Under coincident gauge condition for the non-metricity tensor (2) reduced to
$$ Q_{\beta\mu\tau} = \partial_{\beta}\; g_{\mu\tau}. $$
For modeling of a viable spherical configuration, consider
\begin{equation}\label{11}
ds^2=-e^{\chi}dt^{2}+e^{\psi}dr^{2}+r^2 d\theta^{2}+r^2 \sin^2\theta{d\phi^2},
\end{equation}
where metric functions are function of only $r$.
The non-metricity scalar for above geometry turns out to be
\begin{equation}\label{12}
Q=-\frac{2 e^{{-}{\psi}} \left(r \chi '+1\right)}{r^2}.
\end{equation}
Throughout the article, the prime notion deals with derivative w.r.t $r$.
The mathematically form of anisotropic form of energy-momentum tensor is defined by
\begin{equation}\label{13}
T_{\beta \chi }=(P_t+\rho) V_{\beta} V_{\chi} +g_{\beta \chi}P_t +(P_r - P_t) S_{\beta}
S_{\chi},
\end{equation}
satisfies the following relations
\begin{eqnarray}\notag\label{14}
V^{\beta}=e^{\frac{-\nu}{2}} \delta^\beta_0,~S^{\beta}= e^{\frac{-\lambda}{2}}
\delta_1^\beta,~V^{\beta}V_{\beta}=-1,~S^{\beta}S_{\beta} =1,~S^{\beta}V_{\beta} = 0.
\end{eqnarray}
By taking into account the equations (\ref{11}) and (\ref{13}), the non-zero components of field equations are
\begin{equation}\label{18}
\rho = \dfrac{1}{8\pi} (\frac{f}{2}-f_Q \left(Q+\frac{1}{r^2}+\frac{e^{-\psi(r)} \left(\psi '(r)+\chi '(r)\right)}{r}\right)),
\end{equation}
\begin{equation}\label{19}
 P_{r} = \dfrac{1}{8\pi} (f_Q \left(Q+\frac{1}{r^2}\right)-\frac{f}{2}),
\end{equation}
\begin{equation}\label{20}
 P_{t} = \dfrac{1}{8\pi} (f_Q \left(\frac{Q}{2}-e^{-\psi} \left(\frac{\chi ''}{2}+\left(\chi '- \psi ' \right) \left(\frac{\chi ' }{4}+\frac{1}{2 r}\right)\right)\right)-\frac{f}{2}),
\end{equation}
\begin{equation}\label{21}
\dfrac{cot\theta}{2} Q^{'}f_{QQ} = 0,
\end{equation}
The integration of the above equation leads to linear form of the function $f(Q)$
\begin{equation}\label{23}
\textit{f} =  \beta + j Q.
\end{equation}
Here $j$ and $\beta$ are integration constants.

\section{Vaidya-Tikekar Model and Matching Conditions}

In this section, we employ the physical viable solution, i.e., VT to the field equations (\ref{18}-\ref{20})
to formulate a solution for compact objects in $f(Q)$ gravity. The VT model is given by
\begin{equation}\label{24}
e^{\chi(r)} =\left(C+D \sqrt{(1-U) \left(b^2-r^2\right)}\right)^2 ,
\end{equation}
\begin{equation}\label{25}
e^{\psi(r)}=\frac{b^2-U r^2}{b^2-r^2},
\end{equation}
where $C, D, b$, and $U$ are the constants.
\begin{figure}
\includegraphics[width = 7.2 cm]{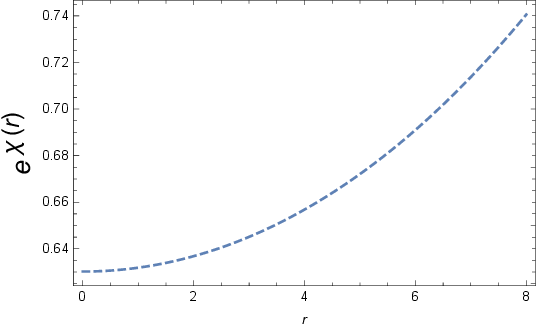} ~~~~~~
\includegraphics[width = 7 cm]{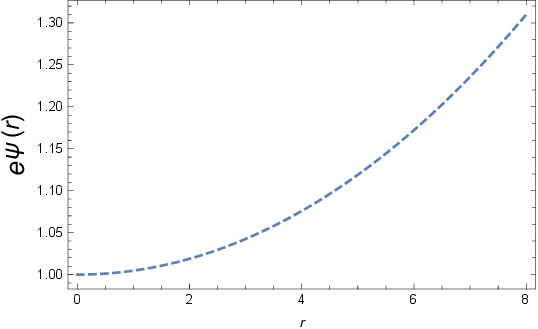}
\caption{Regularity of metric potentials with respect to $r$}
\end{figure}
Consequently, incorporating non-metricity scalar and linear form of $f(Q)$, field equations turns out to be
\begin{equation}\label{26}
\rho = \dfrac{1}{8\pi} \Big[\frac{\beta }{2}+\frac{j  (U-1) \left(3 b^2-U r^2\right)}{\left(b^2-U r^2\right)^2}\Big],
\end{equation}
\begin{equation}\label{27}
\begin{aligned}
P_{r} = \dfrac{1}{8\pi} \Big[(\frac {D \sqrt{(1-U) \left(b^2-r^2\right)} \left(-2 j  (U-3)+\beta
U r^2-\beta  b^2\right)}{2 \left(b^2-U r^2\right) \left(C+D \sqrt{(1-U) \left(b^2-r^2\right)}\right)})\\
+ (\frac{C \left(-2 j  (U-1)+\beta  U r^2-\beta  b^2\right)}{2 \left(b^2-U r^2\right)
\left(C+D \sqrt{(1-U) \left(b^2-r^2\right)}\right)})\Big] ,
\end{aligned}
\end{equation}
\begin{equation}\label{28}
\begin{aligned}
P_{t} = \dfrac{1}{8\pi \Big(2 \left(b^2-U r^2\right)^2 \left(C+D
\sqrt{(1-U) \left(b^2-r^2\right)}\right)\Big)}\times \\
\Big[-\left(\left(C \left(\beta  U^2 r^4+2 b^2 \left(j  (U-1)-
\beta  U r^2\right)+\beta  b^4\right)\right.\right. \\ 
+D \sqrt{(1-U) \left(b^2-r^2\right)} \left(2 b^2 \left(j  (U-3)-\beta  U r^2\right)+
U r^2 \left(2 j +\beta  U r^2\right)+\beta  b^4\right)\Big].
\end{aligned}
\end{equation}
Matching conditions plays a pivotal role for obtained solution viability. Here, we used
Schwarzschild solution for matching and obtaining constants involved in model at the boundary
surface where pressure vanishes. Consider the exterior solution
\begin{equation}\label{30}
ds^2=-\Big(1-\frac{2M}{r}\Big)dt^2+
\Big(1-\frac{2M}{r}\Big)^{-1}dr^2+r^2
d\theta^{2}+r^2 \sin^2\theta{d\phi^2}.
\end{equation}
Using these boundary conditions $r = R, g_{tt}^{-} = g_{tt}^{+}$ and $g_{rr}^{-} = g_{rr}^{+}$,
we have
\begin{equation}\label{31}
e^{\chi}=e^{-\psi}=\Big(1-\frac{2M}{R}\Big), ~~ m(R)=M, ~ P_r(R)=0,
\end{equation}
and it yields the following relations
\begin{equation}\label{32}
 e^{\chi}=\left(C+D \sqrt{(1-U) \left(b^2-r^2\right)}\right)^2=\Big(1-\frac{2M}{R}\Big),
\end{equation}
\begin{equation}\label{33}
e^{\psi} =\frac{b^2-U r^2}{b^2-r^2}=\Big(1-\frac{2M}{R}\Big)^{-1},
\end{equation}
consequently,
\begin{equation}\label{34}
-\frac{2 D (1-U) r}{\sqrt{(1-U) \left(b^2-r^2\right)} \left(C+D \sqrt{(1-U)
\left(b^2-r^2\right)}\right)} = \dfrac{2M}{R^{2}}.
\end{equation}
Additionally, $ P_{r} (R) = 0$ provides
\begin{equation}\label{35}
\begin{aligned}
\dfrac{1}{8\pi} \Big[(\frac {D \sqrt{(1-U) \left(b^2-r^2\right)} \left(-2 j  (U-3)+\beta
U r^2-\beta  b^2\right)}{2 \left(b^2-U r^2\right) \left(C+D \sqrt{(1-U) \left(b^2-r^2\right)}\right)})\\
+ (\frac{C \left(-2 j  (U-1)+\beta  U r^2-\beta  b^2\right)}{2 \left(b^2-U r^2\right)
\left(C+D \sqrt{(1-U) \left(b^2-r^2\right)}\right)})\Big]=0.
\end{aligned}
\end{equation}
From (\ref{32}) to (\ref{35}), we obtain the subsequent
values of constants in terms of model parameter,
\begin{equation}\label{36}
C=\frac{(3-U)}{2}  \sqrt{\frac{R-2 M}{R}},
\end{equation}
\begin{equation}\label{37}
D=-\sqrt{\frac{M}{2 R^3}},
\end{equation}
\begin{equation}\label{38}
b = R \sqrt{\frac{2 U M - U R + R}{2 M}},
\end{equation}
\begin{equation}\label{39}
\beta =-\frac{2 j  \left(C (U-1)+D (U-3) \sqrt{(1-U) \left(b^2-R^2\right)}
\right)}{\left(b^2-U R^2\right) \left(C+D \sqrt{(1-U) \left(b^2-R^2\right)}\right)}.
\end{equation}

\section{Physical Analysis and Validity by LMC X-4}

This section deals with the physical analysis of the formulated model via
Vaidya-Tikekar ansatz and its validation through pulsar LMC X-4. For this,
we have to apply some criterion like regularity of metric potentials, physical
parameters as well as some criterion for its stability guarantee.

\subsection{Metric Potential}

Examining the behaviour of the metric functions at the central point, i.e.,
at $r=0$, is crucial when examining how compact things behaves.
Understanding the existence of material and structural singularities in the compact
stellar structure is a key feature for viability of model
\begin{equation}\label{40}
e^{\psi(r)}\mid\; _{r=0} = 1,
\end{equation}
\begin{equation}\label{41}
e^{\chi(r)}\mid\; _{r=0} = (C+b D \sqrt{(1-U)} )^2,
\end{equation}
shows metric components are free from singularities.

\subsection{The Characteristics of Pressure and Density }

The density and pressure at the center turns out to be
\begin{equation}\label{42}
{\rho(r)}\mid _{r=0}= \frac{6 j  (U-1)+\beta  L^2}{4  L^2},
\end{equation}
\begin{figure}
\centering
\includegraphics[width = 7 cm]{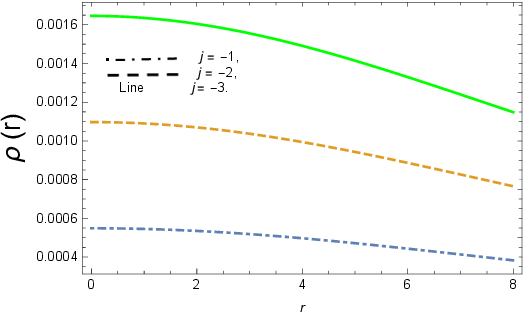}
\caption{The variation of density with respect to $r$}
\end{figure}
\begin{equation}\label{43}
{P_{r}(r)}\mid _{r=0}=-\frac{2 j  C (U-1)+\beta  C b^2+D \sqrt{(1-U) b^2} \left(2 j  (U-3)+
\beta  b^2\right)}{4 b^2 \left(C+D \sqrt{(1-U) L^2}\right)}.
\end{equation}
Figure (\textbf{2} and \textbf{3}), i.e., the behavior of matter parameters have a
decreasing graph from center to the boundary surface, particulary the radial pressure
vanishes at the boundary surface indicate about the compactness of the stellar model.
\begin{figure}
\centering
\includegraphics[width = 6.5cm]{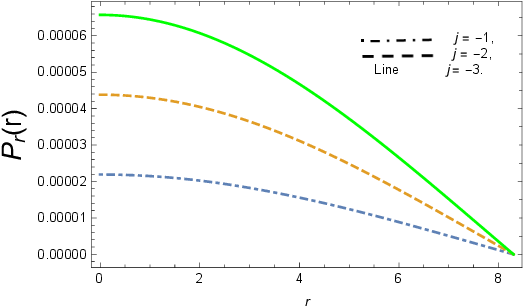} ~~~
\includegraphics[width = 6.5cm]{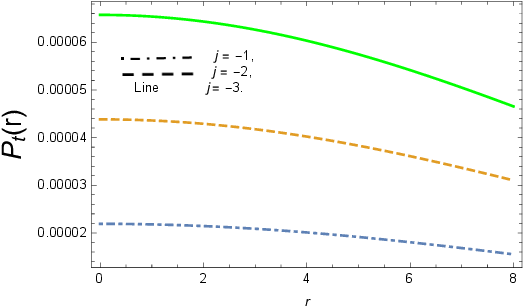}
\caption{The variation of pressure stresses with respect to $r$}
\end{figure}
The gradients of the density and pressure stress from (\ref{26})-(\ref{28}) have the
following equations
\begin{equation}\label{44}
\dfrac{d\rho}{dr}=\dfrac{1}{8 \pi}\Bigg[-\frac{2 j  (U-1) U r \left(U r^2-5 b^2\right)}
{\left(b^2-U r^2\right)^3}\Bigg],
\end{equation}
\begin{equation}\label{45}
\begin{aligned}
\dfrac{dP_{r}}{dr}=\dfrac{1}{8 \pi}\Bigg[\dfrac{2r\left(D^2 U \left(U^2-4 U+3\right)
\left(b^2-r^2\right)^2-C^2 (U-1) U \left(b^2-r^2\right)\right.}{\left(b^2-r^2\right)
\left(b^2-U r^2\right)^2 \left(C+D \sqrt{(1-U) \left(b^2-r^2\right)}\right)^2}\\
- \dfrac{\left.C D \sqrt{(1-U) \left(b^2-r^2\right)} \left(\left(2 U^2-4 U+1\right)
b^2+U (3-2 U) r^2\right)\right)j}{\left(b^2-r^2\right) \left(b^2-U r^2\right)^2
\left(C+D \sqrt{(1-U) \left(b^2-r^2\right)}\right)^2}\Bigg],
\end{aligned}
\end{equation}
\begin{figure}
\centering
\includegraphics[width = 7cm]{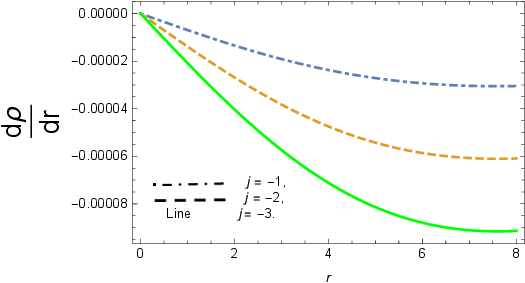}
\caption{The density gradient against radial coordinate $r$}
\end{figure}

\begin{equation}\label{46}
\begin{aligned}
\frac{dP_{t}}{dr} = \frac{1}{8 \pi\Big(\left(b^2-r^2\right) \left(b^2-U r^2\right)^3 \left(C+D \sqrt{(1 - U) 
\left(b^2-r^2\right)}\right)^2\Big)} \times \\
\quad \Bigg[ r\Bigg(2 D^2 (U-1) U \left(b^2-r^2\right)^2
\left((2 U-5) b^2+U r^2\right) \\
\quad -4 C^2 (U-1) U b^2 
 \left(b^2-r^2\right)\quad + C D \sqrt{(1-U) \left(b^2-r^2\right)} \\
 \Big(-2 \left(4 U^2-7 U+1\right) b^4+U^2 r^4 \quad +U (6 U-11) b^2 r^2\Big)\Bigg)j \Bigg].
\end{aligned}
\end{equation}
The plots (\textbf{4} and \textbf{5}) clearly shows the maximum value at the center, i.e., zero
and decreasing further inside the stellar model heading towards boundary surface. 
\begin{figure}
\centering
\includegraphics[width = 6.5cm]{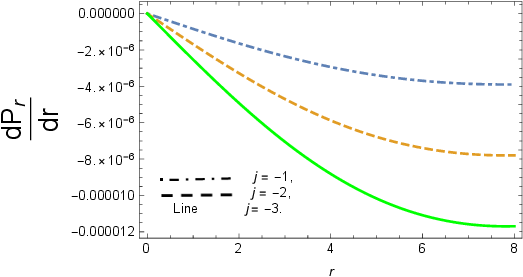} ~~~
\includegraphics[width = 6.5cm]{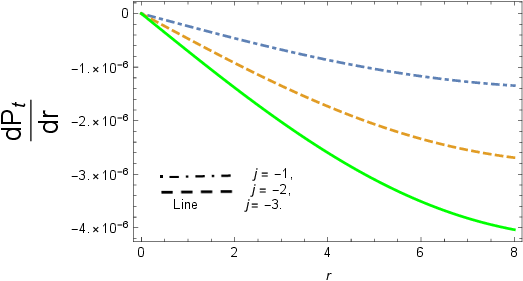}
\caption{The pressure gradient against radial coordinate $r$ }
\end{figure}

\subsection{Anisotropy and Equation of State Parameters}

This section deals with the important parameters anisotropy ($P_{t}-P_{r}$)
and equation of state parameter ($w_{r}, w_{t}$) for model's viability. 
The anisotropy expression in our case is given by
\begin{equation}\label{47}
\Delta = -\frac{j  U r^2 \left(C (U-1)+D (U-2) \sqrt{(1-U) \left(b^2-r^2\right)}\right)}
{8 \pi  \left(b^2-U r^2\right)^2 \left(C+D \sqrt{(U-1) \left(r^2-b^2\right)}\right)}.
\end{equation}
Anisotropic pressure (Fig. \textbf{6}) is orientated outward and $\Delta$ is positive when the tangential
pressure $p_{t}$ is greater than the radial pressure $p_{r}$, and
vice versa indicating an inward pressure.
\begin{figure}
\centering
\includegraphics[width = 7cm]{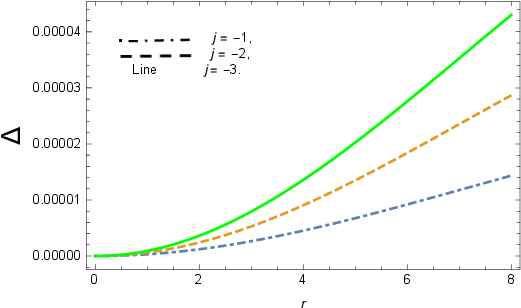}
\caption{Anisotropy factor against radial coordinate $r$}
\end{figure}
The expression for $\omega_{r}$ and $\omega_{t}$ turns out to be
\begin{equation}\label{48}
\begin{aligned}
 \omega_{r} = -\Big[\left(b^2-U r^2\right) \left(D \sqrt{(1-U) \left(b^2-r^2\right)} \left(2 j  (U-3)+\beta
\left(b^2-U r^2\right)\right) \right.+ C \\
\left(2 j (U-1)+\beta  \left(b^2-U r^2\right)\right))\Big]\textbf{/}\\
\Big[ \left(b^2 \left(6 j (U-1)-2 \beta  U r^2\right)+U r^2 \left(\beta  U r^2-2 j (U-1)\right)+\beta  b^4\right)\\
\left(C+D \sqrt{(1-U) \left(b^2-r^2\right)}\right)\Big],
 \end{aligned}
\end{equation}
\begin{equation}\label{49}
\begin{aligned}
\omega_{t} = \frac{-1}{\Big(C+D \sqrt{(1-U) \left(b^2-r^2\right)})(\left.\beta  \left(b^2-U r^2\right)^2-2 j  
 (U-1) \left(U r^2-3 b^2\right)\right)\Big)}  \\
 \Big[(2 j  C (U-1) b^2+\beta  C \left(b^2-U r^2\right)^2+D \sqrt{(1 - U)
 (b^2 - r^2)})\\
 \times (2 j  \left((U-3) b^2+U r^2\right)+\beta  \left(b^2-U r^2\right)^2)\Big].
\end{aligned}
\end{equation}
Both $\omega_{r}$ and $\omega_{t}$ show a monotonic decreasing trend (Fig. \textbf{7})
falling inside the range $ (0, 1) $.
\begin{figure}
\includegraphics[width = 7cm]{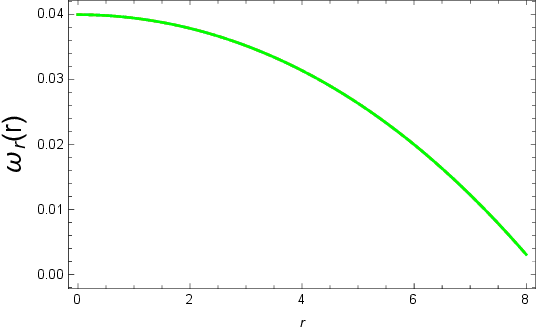} ~~~
\includegraphics[width = 7cm]{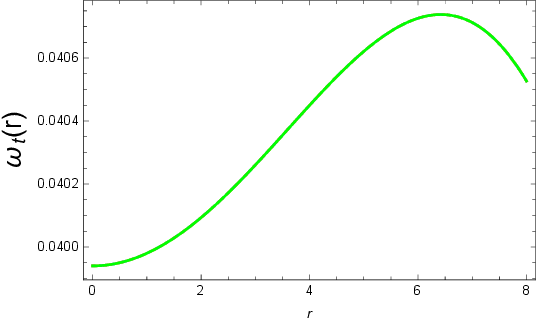}
\caption{The equation of state parameters, $\omega_{r}$ and $\omega_{t}$ vs $r$}
\end{figure}

\subsection{Stability Analysis via Sound Velocity, Adiabatic Index}

The speed of sound can be separated into tangential and
radial components. In general, radial sound speeds are faster 
than tangential sound speeds because of the higher 
gravitational compression in the vicinity of the core. The star's mass,
density, and equation of state all affect the stable ranges for these speeds, which
are generally between $0$ and $1$, i.e., $(0<V_{r}^{2}, V_{t}^{2}<1)$. In order to analyze the
dynamics of compact stars and get insight into their structural stability and
gravitational interactions, it is imperative that these sound speeds be understood.
We obtain the following formulae for the squares of the tangential and radial sound
velocity in our present model,
\begin{equation}\label{50}
\begin{aligned}
V_{r}^{2}= \frac{-1}{\Big((U-1) U \left(b^2-r^2\right) \left(U r^2-5 L^2\right) \left(C+D \sqrt{(1 - U)
 \left(b^2-r^2\right)}\right)^2\Big)} \times \\
 \Big[(b^2-U r^2\left(D^2 U \left(U^2-4 U+3\right) \left(b^2-r^2\right)^2-C^2
(U-1) U \left(b^2-r^2\right)\right.)\\
- \left.C D \sqrt{(1-U) \left(b^2-r^2\right)} \left(\left(2 U^2-4 U+1\right)
 b^2+U (3-2 U) r^2\right)\right))\Big],
\end{aligned}
\end{equation}
\begin{figure}
\includegraphics[width = 7cm]{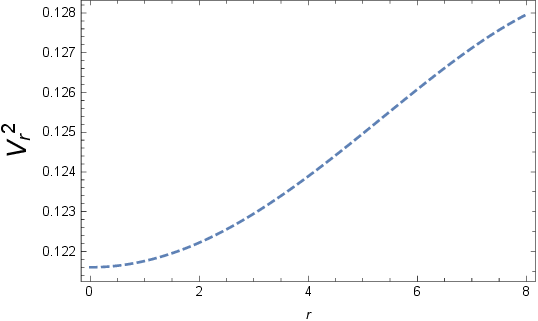} ~~~
\includegraphics[width = 7cm]{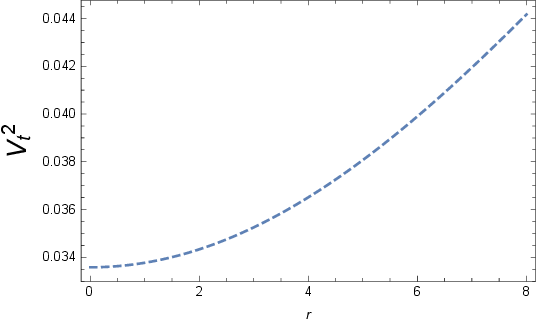}
\caption{Radial and sound speed satisfies the range, i.e., $0 < V_{r}^{2}, V_{t}^{2} < 1$}
\end{figure}
\begin{equation}\label{51}
\begin{aligned}
V_{t}^{2}= \frac{1}{2\Big((U-1) U \left(b^2-r^2\right) \left(U r^2-5 L^2\right) \left(C+D \sqrt{(1 - U)
 \left(b^2-r^2\right)}\right)^2\Big)}\times \\
\Big[(U-1)(4 C^2 U b^2 \left(b^2-r^2\right)-2 D^2 U \left(b^2-r^2\right)^2
\left((2 U-5) b^2+U r^2\right))+ C \\
D \sqrt{(1-U) \left(b^2-r^2\right)} \left(2 \left(4 U^2-7 U+1\right)
b^4-U^2 r^4+U (11-6 U) b^2 r^2\right)\Big].
\end{aligned}
\end{equation}
Figures (\textbf{8} and \textbf{9}) demonstrates that both $ V_{t}^{2}$ and $ V_{r}^{2}$
lies within the stability range.
\begin{figure}
\centering
\includegraphics[width = 7cm]{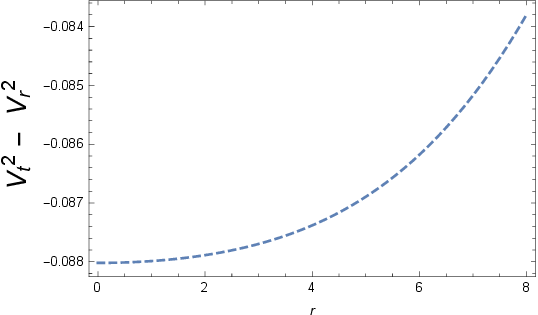}
\caption{The stability factor $V_{t}^{2} - V_{r}^{2}$ against $r$}
\end{figure}
The other important parameter for characterizing the equation of state's stiffness is the
adiabatic index ($\Gamma$) and is essential for determining nature of compact stars, i.e.,
relativistic or non-relativistic as well as its stability. 
It is well established that the value of $\Gamma>\frac{4}{3}$ reveals object stability 
otherwise unstable. The expression for $ \Gamma $ is defined by
\begin{equation}\label{52}
\Gamma = (\dfrac{\rho + p_{r}}{p_{r}}) V_{r}^{2},
\end{equation}
\begin{equation}\label{53}
\begin{aligned}
\Gamma = \frac{A}{B},
\end{aligned}
\end{equation}
where \\
$
A = \Big[(4\left(C^3 (U-1) U b^2 \left(b^2-r^2\right)-D^3 (U-3) U^2 \left(b^2-r^2\right)^3\right.\\
\sqrt{(1-U) \left(b^2-r^2\right)}- C D^2 U \left(b^2-r^2\right)^2 \left(\left(3 U^2-8 U+4\right) b^2+U (3-2 U) r^2\right)+\\
(\left.C^2 D \sqrt{(1-U) \left(b^2-r^2\right)} \left(\left(3 U^2-4 U+1\right) b^4+U^2 r^4+U
 (3-4 U) b^2 r^2\right)\right)j)\Big],$ \\
$ B = \Big [(U (b^2-r^2) \left(U r^2-5 b^2\right) \left(C+D \sqrt{(1-U) \left(b^2-r^2\right)}\right)^2) (C (-2 j \\
(U-1)+\beta  U r^2-\beta  b^2)+D \sqrt{(1-U) \left(b^2-r^2\right)} \left(-2 j  (U-3)+\beta  U r^2-\beta  b^2\right))\Big].$

The Fig. \textbf{10} shows that $\Gamma$ fulfills the stability criterion.
\begin{figure}
\centering
\includegraphics[width = 7cm]{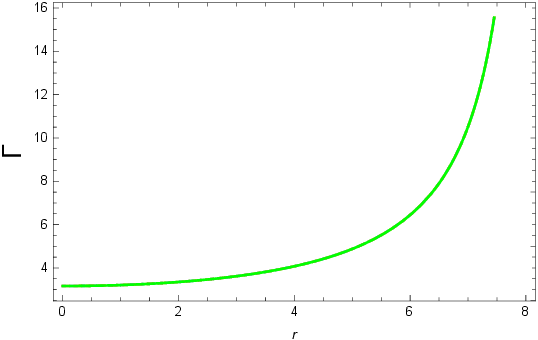}
\caption{For each value of $j=-1,-2,-3$, adiabatic index have value greater than $4/3$}
\end{figure}

\subsection{Compactness and Surface Redshift Parameters}

The following formula can be used to calculate the effective mass \cite{46}
\begin{equation}\label{54}
 M_{eff} = \frac{R}{2}(1- e^{-\psi(R)})\\
 = \frac{(U-1) r^3}{2 U r^2-2 b^2},
\end{equation}
and the compactness parameter is calculated as
\begin{equation}\label{55}
\textbf{\textit{U(r)}} = \dfrac{ M_{eff}}{R} = \frac{(U-1) r^2}{2 U r^2-2 b^2}.
\end{equation}
\begin{figure}
\centering
\includegraphics[width = 6.5cm]{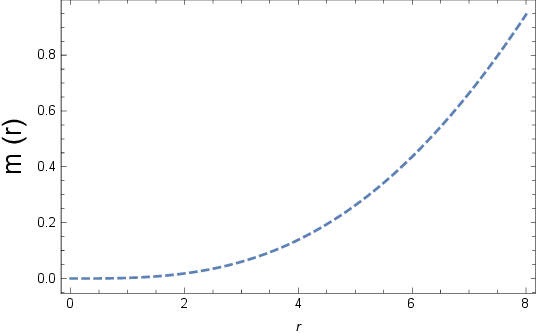} ~~~
\includegraphics[width = 6.5cm]{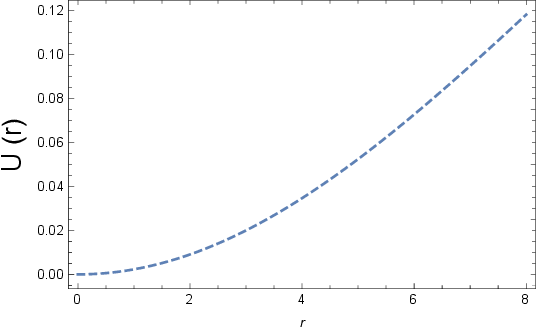}
\caption{Mass function and Compactness against $r$}
\end{figure}
Using compactness, one can write surface redshift parameter as
\begin{equation}\label{56}
z_{s} =\frac{1}{\sqrt {1 - 2 \textbf{\textit{U(r)}}}} - 1.
\end{equation}
Figure (\textbf{11} and \textbf{12}) show that both parameters are well behaved and have a
steady growth. A star's surface redshift
provides important information on the complex physical interactions between its core
components and the equation of state that determines its characteristics. Compact stars
in our particular case satisfy the Buchdahl condition $ u(R) < \dfrac{4}{9}$ \cite{47}.
\begin{figure}
\centering
\includegraphics[width = 7cm]{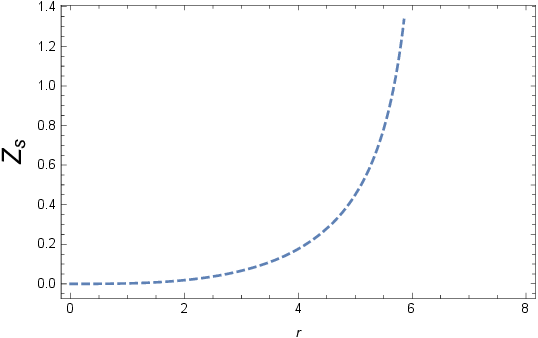}
\caption{Surface Redshift Function against $r$ }
\end{figure}
Moreover, Barraco and Hamity \cite{48} obtained a limit, i.e., $z_{s}\leq 2$
for a physically viable model without cosmological constant. Later, Bohmer and Harko
\cite{49} extended this limit to $z_{s} \leqslant 5$ both with or with out cosmological
constant. Our model satisfies the required criterion. 

\subsection{Energy Conditions}

A set of rules that control the distribution and movement of matter and energy in
space-time are known as energy conditions in GR. They offer
limitations on the distribution of momentum and energy in a specific area of
space-time, known as the energy-momentum tensor.

These conditions include:\\
\textbf{i)} According to the null energy condition (NEC), any contracted energy-momentum
vector with a null vector is non-negative, i.e., $\rho \geqslant 0$.\\
\begin{figure}
\centering
\includegraphics[width = 6.5cm]{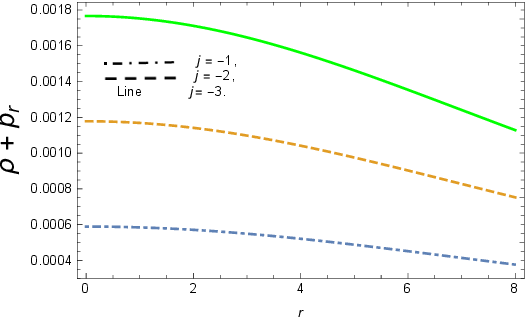} ~~~
\includegraphics[width = 6.5cm]{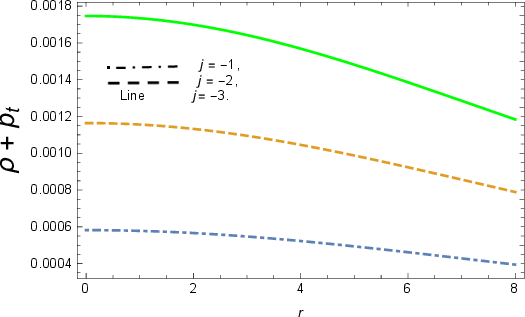}
\caption{WEC against radial coordinate}
\end{figure}
\textbf{ii)} The weak energy condition (WEC), which asserts that the energy-momentum vector
contracted with any timelike vector is non-negative, i.e.,
$ \rho + p_{r} \geqslant 0, \;\;\;\;  \rho + p_{t}  \geqslant 0$.\\
\textbf{iii)} A specific combination of the energy-momentum tensor components must be non-negative,
according to the strong energy condition (SEC), i.e., $\rho + p_{r} + 2p_{t} \geqslant 0$.\\
\begin{figure}
\centering
\includegraphics[width = 7cm]{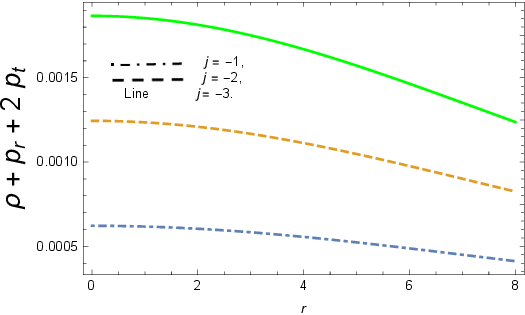}
\caption{SEC against radial coordinate}
\end{figure}
\textbf{iv)} The dominant energy condition (DEC), which requires that the energy-momentum vector
contracted with any future-directed causal vector is non-negative, i.e.,
$ \rho - |p_{r}|  \geqslant 0, \;\;\;\;  \rho - |p_{t}| \geqslant 0$.\\
\begin{figure}
\centering
\includegraphics[width = 6.5cm]{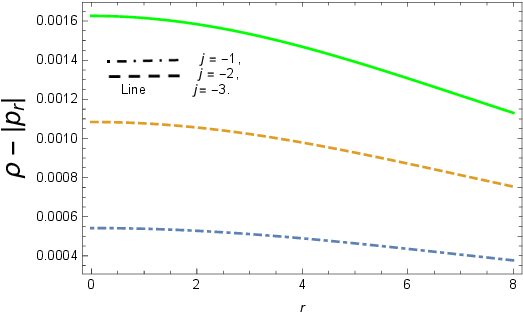} ~~~
\includegraphics[width = 6.5cm]{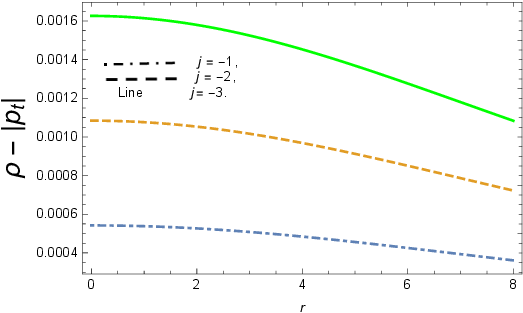}
\caption{DEC against radial coordinate}
\end{figure}
All the conditions are satisfies for out model shown in Figs. (\textbf{13}, \textbf{14} and \textbf{15}).

\subsection{The stability standard outlined by Harrison, Zeldovich, and
Novikov for static configurations}

Researchers like Chandrasekhar \cite{50} and Harrison et al. \cite{51}
employed a variety of techniques to assess the steadiness of stellar formations.
Zeldovich and Novikov \cite{52} then made more straightforward by computations using
the findings of Harrison et al. \cite{51}. Their research states that
$ \dfrac{\partial M}{\partial \rho_{c}} > 0 $ indicates that the model is stable, and
that the inequality is reversed for an unstable model. When $\rho_{c}$ denotes the
central density and the following expression for $M$ is given,
 \begin{equation}\label{61}
 M(\rho_{c}) = \frac{(U-1) R^3 \left(\beta -4 \rho _c\right)}{2 \left(-4 U R^2
 \rho _c+6 j  (U-1)+\beta  U R^2\right)}.
 \end{equation}
Now differentiating this equation partially we get,
\begin{equation}\label{62}
 \dfrac{\partial M}{\partial \rho_{c}} =\frac{12 j  (U-1)^2 R^3}{\left(-4 U R^2
 \rho _c+6 j  (U-1)+\beta  U R^2\right){}^2}.
\end{equation}
\begin{figure}
\includegraphics[width = 7cm]{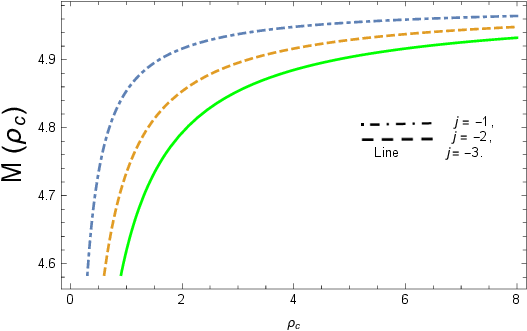} ~~~
\includegraphics[width = 7.7cm]{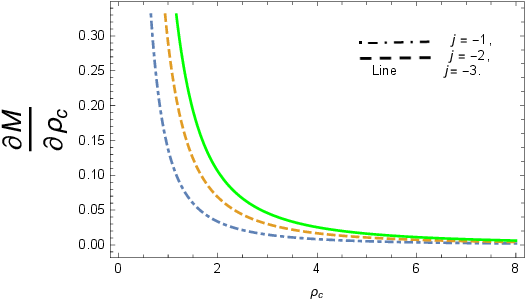}
\caption{Left and right panel of $M(\rho_{c})$ and $\dfrac{\partial M}{\partial\rho_{c}}$, respectively for fixed values of $j=-1,-2,-3$}
\end{figure}
For different values of $j$, it is evident from Figure \textbf{16} that both $M (\rho_{c})$
and $\dfrac{\partial M}{\partial \rho_{c}}$ is positive everywhere inside the stellar interior.

\section{The TOLMAN-OPPENHEIMER-VOLKOFF Equation}

The Tolman-Oppenheimer-Volkoff (TOV) equation is defined by
\begin{equation}\label{63}
-\frac{\chi^\prime(\rho + P_{r})}{2}- \frac{dP_{r}}{dr} + 2\frac{(P_{t}-P_{r})}{r} = 0,
\end{equation}
\begin{figure}
 \centering
\includegraphics[width = 7cm]{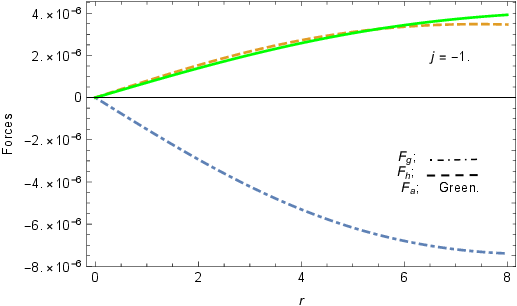}
\caption{Behavior of radial forces in TOV equation for $j=-1$}
\end{figure}
Models of physical reality should be able to endure stability tests using gravitational
force $(F_{g})$, hydrostatic force $(F_{h})$, and anisotropic force $(F_{a})$, according
to Tolman \cite{53}, Oppenheimer, and Volkoff \cite{54}. For the system to remain in
equilibrium, the net effect of all these forces must be zero, mathematically,
$F_{g} + F_{h} + F_{a} = 0$. The following equations are given to represent these forces,
\begin{equation}\label{64}
F_{g} = \frac{j  D (U-1) r \left(D U \left(b^2-r^2\right)^2-C b^2 \sqrt{(1-U)
\left(b^2-r^2\right)}\right)}{4 \pi  \left(r^2-b^2\right) \left(b^2-U r^2\right)^2
\left(C+D \sqrt{(U-1) \left(r^2-b^2\right)}\right)^2},
\end{equation}
\begin{equation}\label{65}
\begin{aligned}
F_{h}= - \dfrac{1}{8 \pi}\Bigg[(\dfrac{2r\left(D^2 U \left(U^2-4 U+3\right)
\left(b^2-r^2\right)^2-C^2 (U-1) U \left(b^2-r^2\right)\right.}{\left(b^2-r^2\right)
\left(b^2-U r^2\right)^2 \left(C+D \sqrt{(1-U) \left(b^2-r^2\right)}\right)^2})\\
- (\dfrac{\left.C D \sqrt{(1-U) \left(b^2-r^2\right)} \left(\left(2 U^2-4 U+1\right)
b^2+U (3-2 U) r^2\right)\right)j}{\left(b^2-r^2\right) \left(b^2-U r^2\right)^2
\left(C+D \sqrt{(1-U) \left(b^2-r^2\right)}\right)^2})\Bigg],
\end{aligned}
\end{equation}
\begin{figure}
 \centering
\includegraphics[width = 7cm]{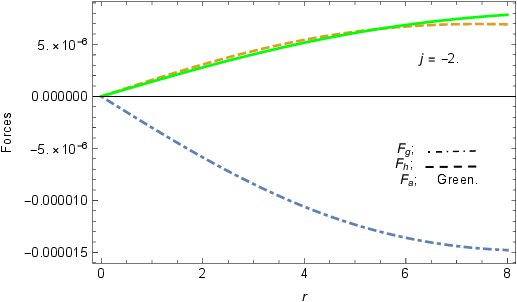}
\caption{Behavior of radial forces in TOV equation for $j=-2$}
\end{figure}
\begin{equation}\label{66}
F_{a} = -\frac{j  U r \left(C (U-1)+D (U-2) \sqrt{(1-U) \left(b^2-r^2\right)}\right)}
{4 \pi  \left(b^2-U r^2\right)^2 \left(C+D \sqrt{(U-1) \left(r^2-b^2\right)}\right)}.
\end{equation}
Plots (\textbf{17}, \textbf{18} and \textbf{19}) showed that the combined influence of the
gravitational, hydrostatic, and anisotropic forces significantly supports the condition of
equilibrium inside our system.
\begin{figure}
 \centering
\includegraphics[width = 7cm]{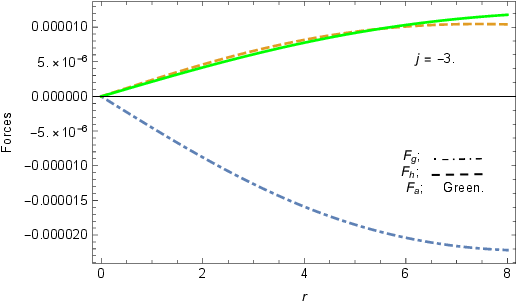}
\caption{Behavior of radial forces in TOV equation for $j=-3$}
\end{figure}

\section{Conclusion}

We investigate solutions pertaining to Compact stars under the formalism of
$f (Q)$ theory, a development of symmetric teleparallel theory. We integrate random
functions of the non-metricity scalar $Q$ using a coincident gauge. We make use of the
VT potential to construct analytical models for compact stars in $f (Q)$ gravity.
The matter components are determined explicitly by using VT functions in conjunction
with a linear formulation of $f (Q)$. Our assumption is that the interior of the star is a static,
spherically symmetric source. The observed mass and radius values for the compact star LMC X-4,
which has a mass of $(1.04 \pm 0.09)$ solar masses and a radius of $(8.301 \pm 0.2)$ kilometres,
are also used to calculate the unknown constants of the VT model. In the context of $f (Q)$ gravity, 
LMC X-4 serves as an ideal astrophysical test bed for exploring the effects of this framework 
on compact objects and their environments. Neutron stars in such systems provide strong-field 
and high-density conditions where deviations from general relativity may manifest, 
making LMC X-4 particularly relevant for investigating how $f (Q)$ gravity influences the 
neutron star's mass-radius relationship, equation of state, internal structure, 
pulsar behavior, and energy emission processes.

With $ e^{\psi(r)}|_{r=0} = 1 $ and $ e^{\chi (r)}|_ {r=0} = (C+b D \sqrt{(1-U)})^2 $,
graphical analysis demonstrates that the metric potentials fulfill the necessary criteria in
the centre, confirming solutions free from geometric and material singularities. $e^{\psi}$ and
$e^{\chi}$ are metric potentials that steadily increase in our model. Pressures and matter density
both decrease monotonically with radius $'r'$, staying positive for the duration of the star. The
inner regions of the star have increasing pressure and density magnitudes as $j$ lowers. The
pressure and density gradients are initially negative, reaching their greatest values in the centre.
$\omega{r}$ and $\omega{t}$ are not significantly affected by differences in $j$ values. For
different $j$ values, graphs illustrating the connections between mass and radius, compactness,
and surface redshift fall within predicted ranges. In the end, hydrostatic and anisotropic
forces are balanced by gravitational forces, guaranteeing our incredibly dense compact star's steady
equilibrium. In conclusion, our research shows that the Vaidya-Tikekar metric accurately captures the
properties of a compact star that is stable within a population of stars that are comparable to one
another and free from singularities.


\vspace{0.1cm}


\begin{thebibliography}{40}

\bibitem{1} Aldrovandi, Ruben, and Jose G. Pereira. Teleparallel gravity: an introduction.
Vol. 173. Springer Science \& Business Media, 2012.
\bibitem{2} Maluf, J. W. (2013). The teleparallel equivalent of general relativity.
Annalen der Physik, 525(5), 339-357.
\bibitem{3} Dialektopoulos, K. F., Koivisto, T. S., \& Capozziello, S. (2019). Noether
symmetries in symmetric teleparallel cosmology. The European Physical Journal C, 79, 1-12.
\bibitem{4} Barros, B. J., Barreiro, T., Koivisto, T., \& Nunes, N. J. (2020). Testing $F(Q)$
gravity with redshift space distortions. Physics of the Dark Universe, 30, 100616.
\bibitem{5} Jiménez, J. B., Heisenberg, L., Koivisto, T., \& Pekar, S. (2020). Cosmology in $F(Q)$
geometry. Physical Review D, 101(10), 103507.
\bibitem{6} Bajardi, F., Capozziello, S., \& Vernieri, D. (2020). Non-local curvature and Gauss Bonnet
cosmologies by Noether symmetries. The European Physical Journal Plus, 135(12), 1-15.
\bibitem{7} Ayuso, I., Lazkoz, R., \& Salzano, V. (2021). Observational constraints on cosmological
solutions of $f (Q)$ theories. Physical review d, 103(6), 063505..
\bibitem{8} Flathmann, K., \& Hohmann, M. (2021). Post-Newtonian limit of generalized symmetric
teleparallel gravity. Physical Review D, 103(4), 044030.
\bibitem{9} Khyllep, W., Paliathanasis, A., \& Dutta, J. (2021). Cosmological solutions and growth
index of matter perturbations in $f (Q)$ gravity. Physical Review D, 103(10), 103521.
\bibitem{10}D'Ambrosio, F., Garg, M., \& Heisenberg, L. (2020). Non-linear extension of non-metricity
scalar for MOND. Physics Letters B, 811, 135970.

\bibitem{11} Adak, M., \& Sert, \"O. (2005). A solution to symmetric teleparallel gravity. Turkish
Journal of Physics, 29(1), 1-7.
\bibitem{12} Adak, M., Kalay, M., \& Sert, \"O. (2006). Lagrange formulation of the symmetric
teleparallel gravity. International Journal of Modern Physics D, 15(05), 619-634.
\bibitem{13} Adak, M., Sert, \"O., Kalay, M., \& Sari, M. (2013). Symmetric teleparallel gravity:
some exact solutions and spinor couplings. International Journal of Modern Physics A, 28(32), 1350167.
\bibitem{14} Mol, I. (2017). The non-metricity formulation of general relativity. Advances in
Applied Clifford Algebras, 27, 2607-2638.
\bibitem{15} Jim\'enez, J. B., Heisenberg, L., \& Koivisto, T. S. (2018). Teleparallel palatini theories.
Journal of Cosmology and Astroparticle Physics, 2018(08), 039.
\bibitem{16} Gakis, V., Krs\v{s}\v{s}\'ak, M., Said, J. L., \& Saridakis, E. N. (2020). Conformal gravity and
transformations in the symmetric teleparallel framework. Physical Review D, 101(6), 064024.
\bibitem{17} Zhao, D. (2022). Covariant formulation of $f(Q)$ theory. The European Physical Journal C,
82(4), 303.
\bibitem{18} Lazkoz, R., Lobo, F. S., Ortiz-Ba\`nos, M., \& Salzano, V. (2019). Observational constraints
of $f(Q)$ gravity. Physical Review D, 100(10), 104027.
\bibitem{19} Mandal, S., Sahoo, P. K., \& Santos, J. R. (2020). Energy conditions in $f(Q)$ gravity.
Physical Review D, 102(2), 024057.
\bibitem{20} Capozziello, S., \& Shokri, M. (2022). Slow-roll inflation in $f(Q)$ non-metric gravity.
Physics of the Dark Universe, 37, 101113.
\bibitem{21} Capozziello, S., \& D'Agostino, R. (2022). Model-independent reconstruction of $f(Q)$
non-metric gravity. Physics Letters B, 832, 137229.

\bibitem{22} Jiménez, J. B., Heisenberg, L., Koivisto, T., \& Pekar, S. (2020).
Cosmology in $f (Q)$ geometry. Physical Review D, 101(10), 103507.
\bibitem{23} Jiménez, J. B., Heisenberg, L., \& Koivisto, T. (2018). Coincident
general relativity. Physical Review D, 98(4), 044048.
\bibitem{24} Harko, T., Koivisto, T. S., Lobo, F. S., Olmo, G. J., \& Rubiera-Garcia, D.
(2018). Coupling matter in modified $f (Q)$ gravity. Physical Review D, 98(8), 084043.
\bibitem{25} Solanki R., Pacif J. K. S., Parida A., \& Sahoo K. P. (2021). Cosmic
acceleration with bulk viscosity in modified $f (Q)$ gravity. Phys. Dark Universe 32, 100820.
\bibitem{26} Junior, J. T. S., \& Rodrigues, M. E. (2023). Coincident $f (Q)$ gravity:
black holes, regular black holes, and black bounces. The European Physical Journal C, 83(6), 475.
\bibitem{27} Parsaei, F., Rastgoo, S., \& Sahoo, P. K. (2022). Wormhole in $f (Q)$ gravity.
The European Physical Journal Plus, 137(9), 1-16.
\bibitem{28} Banerjee, A., Pradhan, A., Tangphati, T., \& Rahaman, F. (2021). Wormhole geometries
in $f (Q)$ gravity and the energy conditions. The European Physical Journal C, 81, 1-7.
\bibitem{29} Mustafa, G., Hassan, Z., Moraes, P. H. R. S., \& Sahoo, P. K. (2021). Wormhole solutions
in symmetric teleparallel gravity. Physics Letters B, 821, 136612.
\bibitem{30} De, A., \& How, L. T. (2022). Comment on “Energy conditions in $f (Q)$ gravity”.
Physical Review D, 106(4), 048501.
\bibitem{31} Lin, R. H., \& Zhai, X. H. (2021). Spherically symmetric configuration in $f (Q)$
gravity. Physical Review D, 103(12), 124001.
\bibitem{32} Hassan, Z., Ghosh, S., Sahoo, P. K., \& Rao, V. S. H. (2023). GUP corrected Casimir
wormholes in $f (Q)$ gravity. General Relativity and Gravitation, 55(8), 90.
\bibitem{33} Mandal, S., Sahoo, P. K., \& Santos, J. R. (2020). Energy conditions in $f (Q)$
gravity. Physical Review D, 102(2), 024057.
\bibitem{34} Heisenberg, L. (2024). Review on $f (Q)$ gravity. Physics Reports, 1066, 1-78.
\bibitem{35} Arora, S., Santos, J. R. L., \& Sahoo, P. K. (2021). Constraining $f (Q,T)$ gravity
from energy conditions. Physics of the Dark Universe, 31, 100790.
\bibitem{36} Xu, Y., Li, G., Harko, T., \& Liang, S. D. (2019). $f (Q,T)$ gravity. The European
Physical Journal C, 79, 1-19.
\bibitem{37} Adeel, M., Zeeshan Gul, M., Rani, S., \& Jawad, A. (2023). Physical analysis of anisotropic
compact stars in $f (Q)$ gravity. Modern Physics Letters A, 38(34n35), 2350152.
\bibitem{38} Bhar, P., Malik, A., \& Almas, A. (2024). Impact of $f (Q)$ gravity on anisotropic
compact star model and stability analysis. Chinese Journal of Physics, 88, 839-856.
\bibitem{39} Maurya, S. K., Errehymy, A., Vîlcu, G. E., Alrebdi, H. I., Nisar, K. S., \& Abdel-Aty, A. H.
(2024). Anisotropic compact star in linear $f (Q)$-action. Classical and Quantum Gravity, 41(11), 115009.
\bibitem{40} Gogoi, D. J., Övgün, A., \& Koussour, M. (2023). Quasinormal modes of black holes in $f (Q)$
gravity. The European Physical Journal C, 83(8), 700.
\bibitem{41} Chanda, A., \& Paul, B. C. (2022). Evolution of primordial black holes in $f (Q)$ gravity
with non-linear equation of state. The European Physical Journal C, 82(7), 1-11.
\bibitem{42} Awais, M., \& Azam, M. (2025). Modeling of charged anisotropic compact star with Tolman-IV ansatz. 
International Journal of Geometric Methods in Modern Physics, 2540009.
\bibitem{43} Calzá, M., \& Sebastiani, L. (2023). A class of static spherically symmetric solutions in $f (Q)$-
gravity. The European Physical Journal C, 83(3), 1-9.
\bibitem{44} Chanda, A., \& Paul, B. C. (2022). Evolution of primordial black holes in $f (Q)$ gravity with
non-linear equation of state. The European Physical Journal C, 82(7), 1-11.
\bibitem{45}  Jiménez, J. B., Heisenberg, L., \& Koivisto, T. (2018). Coincident general relativity.
Physical Review D, 98(4), 044048.
\bibitem{46} Goncalves, V. P., \& Lazzari, L. (2020). Electrically charged strange stars with an interacting
quark matter equation of state. Physical Review D, 102(3), 034031.
\bibitem{47} Buchdahl, H. A. (1959). General relativistic fluid spheres. Physical Review, 116(4), 1027.
\bibitem{48} Barraco, D., \& Hamity, V. H. (2002). Maximum mass of a spherically symmetric isotropic star.
Physical Review D, 65(12), 124028.
\bibitem{49} Böhmer, C. G., \& Harko, T. (2006). Bounds on the basic physical parameters for anisotropic
compact general relativistic objects. Classical and Quantum Gravity, 23(22), 6479.
\bibitem{50} Chandrasekhar, S. (1964). Dynamical instability of gaseous masses approaching the
Schwarzschild limit in general relativity. Physical Review Letters, 12(4), 114.
\bibitem{51} Harrison, B. K., Thorne, K. S., Wakano, M., \& Wheeler, J. A. (1965). Gravitation theory
and gravitational collapse. Gravitation Theory and Gravitational Collapse.
\bibitem{52} Zeldovich, Y. B., \& Novikov, I. D. (1971). Relativistic astrophysics. Vol. 1:
Stars and relativity. Chicago: University of Chicago Press.
\bibitem{53} Tolman, R. C. (1939). Static solutions of Einstein's field equations for spheres of fluid.
Physical Review, 55(4), 364.
\bibitem{54} Oppenheimer, J. R., \& Volkoff, G. M. (1939). On massive neutron cores. Physical Review, 55(4), 374.

\end{thebibliography}
\end{document}